\documentclass[journal,draftcls,onecolumn,12pt,twoside]{IEEEtranTCOM}
\normalsize
\ifCLASSINFOpdf
\else
\fi
\usepackage{subcaption}
\usepackage{stfloats}
\usepackage{amsmath}
\usepackage{amsthm}
\usepackage{graphicx}
\usepackage{tipa}
\usepackage{changes}
\usepackage{mathtools}
\usepackage{amssymb}

\ifCLASSOPTIONcompsoc
\else

\DeclareFontFamily{U}{mathx}{\hyphenchar\font45}
\DeclareFontShape{U}{mathx}{m}{n}{
	<5> <6> <7> <8> <9> <10>
	<10.95> <12> <14.4> <17.28> <20.74> <24.88>
	mathx10
}{}

\begin{document}
\title{Resource Allocation in MIMO setup}
\author{Felix Ma Yun, Jordan Nabi, Mitra Hassani}

\maketitle
\begin{abstract}
In a multi-input multi-output (MIMO) setup, where one side of the link comprises linear antenna array, data can be transmitted over the direction of incident rays.  Channel capacity for this setup is studied in this paper. We define two different setups; one when the energy is constant and equal over all rays, one when available energy is evenly distributed over rays. For the latter, we show that there is an upper bound for channel capacity, regardless of the number of rays and antennas. Also, we have compared this setup with the legacy single-input single-output (SISO) AWGN channel.
\end{abstract}

\section{Introduction}

Multi-input multi-output (MIMO) communication techniques have been widely used in the recent decades. This method has been unbeatable in multi-path environment. Using multi-element antenna rays at both receiver and transmitter exploits spatial features in addition to time and frequency division \cite{hassan1996cryptographic}. In a paper published in 1998 [1] it was shown that channel capacity increases linearly with the number of antennas for every 3 dB increase in SNR under the assumption of independent Rayleigh fading paths between antenna elements. Many other papers have investigated linear relationship between MIMO capacity and the number of antennas under some conditions. \cite{hamidi2019systems} [2] [3]. It is worth mentioning to say that the conditions defined in these papers could not be fully satisfied in all environment. \\
Another limitation on MIMO channel capacity arises from the correlation between individual sub-channels of the channel matrix [1,4,5]. We know that for any increase in correlation coefficient, channel capacity would decrease correspondingly, and in a special case, when correlation coefficient is equal to unity, there is no advantage using MIMO channel. How this correlation affects MIMO channel capacity has been studied in [4]. However, the method used in this paper, does not investigate the effect of correlation explicitly, meaning that a relation between channel capacity versus correlation have not been calculated.\\
Considering a multi-element antenna (MEA) system, it has been shown that as n=min(nT,nR) goes to infinity, for a given fixed transmit power, if the fading between pairs of transmit-receive antenna elements are independent and Rayleigh, the average channel capacity divided by n approaches a constant number determined by SNR [1].\\
In this paper, to get around the limitations caused by the correlation between pairs of transmit-receive antenna elements, we want to transmit data over direction of rays. In this setup, the number of receivers (nR) and transmitters (nT) would be the number of antennas and number of distinguished-direction rays. What determines the number of transmitters in this setup is the number of rays with different incident directions. Also, we assume that the total available transmitting power is constant. With having this in mind, we define two different setups based on the way that energy is allocated to each ray.

\section{Definitions and Assumptions}

\subsection{Notation and System Model}
In this paper we use $x^{\star}$, $x^{\prime}$ and $x^{\dag}$ to show conjugate, transpose and conjugate transpose of a vector x, respectively. Also, $n_{R}$ and $n_{T}$ are the number of receivers and transmitters respectively. 
We assume that communication is performed while channel can be regarded as essentially fixed.\\
 Let the signal carried over the I’th ray denoted by $s^{i}(t)$ and the signal received by the $j^{th}$ antenna element is denoted by $r^{j}(t)$, and v(t) is additive white Gaussian noise (AWGN). The impulse response connecting the input of the $l^{th}$ channel to output of the $m^{th}$ channel is denoted by $h^{m,l}(t)$. If the communication bandwidth is narrow enough that the channel frequency response can be treated as flat across frequency and therefore we have: 
\begin{align}
r_{\tau}=Hs_{\tau}+v_{\tau}
\end{align} 
where $\tau$ is the discrete-time index. 
A real Gaussian random variable with mean $\mu$ and variance $\sigma^2$ is denoted as $N(\mu,\sigma^2)$. Circularly symmetric complex Gaussian random variable $z$ denoted by $z~\sim(0,\sigma^2)$, is a random variable $z=x+iy$ in which $x$ and $y$ are i.i.d. with $x$ and $y ~N\sim(0,\sigma^2/2)$.

\subsection{Determine matrix H}
In the setup we defined, $n_{R}$ and $n_{T}$ are indeed the number of antennas and rays, respectively. We assume that the information we want to transmit is carried over the rays coming from different directions. Let us say we want to have data transmitted over  $n_{T}$ rays. Then we divide up $(0, 180^{\circ})$ into $n_{T}$ equally spaced angels. In this case, the corresponding angels would be \big( $0^{\circ}, \frac{180^{\circ}}{n_{T}}, 2\times \frac{180^{\circ}}{n_{T}},...,(n_{T}-1)\times\frac{180^{\circ}}{n_{T}} $  \big).
 If we assume that we use linear antenna with $n$ elements at the receiver, then the gain induced at $k^{th}$ antenna knowing that the incident angle is  $\theta$  would be: \\

\begin{align}
e^{jkdcos(\theta)}
\end{align} 
In this formula, we assume that antennas are equally spaced with inter-element distance of d, and $k=\frac{2\pi}{\lambda}$.\\
To make the matrix $H$, each column represents the rays received by a single elements, to be more specific, $H$ is a $n_{R}\times n_{T}$ matrix with $h_{i,j}$ equals to the gain received by the $i^{th}$ antenna from the $j^{th}$ ray. To constitute the first column, which corresponds to the first ray over $n_{R}$ antennas, we have:  \\
$
H_{:,1}=
\begin{pmatrix*}[c]
1 \\
e^{jkdcos(0)} \\
e^{jk2dcos(0)}\\
\\
.\\
.\\
.\\
e^{jk (n_{R}-1)dcos(0)}\\
\end{pmatrix*}
$
\\
For the next column, which is corresponding to the second ray, the incident angle is $\frac{180^{\circ}}{n_{T}}$ and therefore we have:\\
$
H_{:,2}=
\begin{pmatrix*}[c]
1 \\
e^{jkdcos(\frac{180^{\circ}}{n_{T}})} \\
e^{jk2dcos(\frac{180^{\circ}}{n_{T}})}\\
\\
.\\
.\\
.\\
e^{jk (n_{R}-1)dcos(\frac{180^{\circ}}{n_{T}})}\\
\end{pmatrix*}
$
\\
Hence, we can make the entire H as follows:\\
$
H=
\begin{pmatrix*}[c]
1 & 1 & ... & 1 \\
e^{jkdcos(0)} & e^{jkdcos(\frac{180^{\circ}}{n_{T}})} & . . . & e^{jkdcos((n_{T}-1)\times\frac{180^{\circ}}{n_{T}})}\\
e^{jk2dcos(0)} & e^{jk2dcos(\frac{180^{\circ}}{n_{T}})} & . . . & e^{jk2dcos((n_{T}-1)\times\frac{180^{\circ}}{n_{T}})}\\
.\\
.\\
.\\
e^{jk (n_{R}-1)dcos(0)} & e^{jk (n_{R}-1)dcos(\frac{180^{\circ}}{n_{T}})} & . . . & e^{jk (n_{R}-1)dcos((n_{T}-1)\times\frac{180^{\circ}}{n_{T}})}\\
\end{pmatrix*}
$

\section{MIMO Channel Capacity}
In the case where the origin does not know the channel condition, the best strategy is to transmit the transmitting antennas at equal power. In this case, the MIMO channel capacity is calculated as follows:\\

\begin{align}
C= \log_2 det(I_{n_{R}}+\frac{P}{\sigma^2  n_{T}} HH^{\dag}) \ \ \ bit/s/Hz
\end{align} \\
where $I_{n_{R}}$ is the $n_{R}\times n_{R}$ unit matrix and $\sigma^2$ is the noise power.

In the setup we have, however, we assume that each antenna contributes equally to the pattern of whole array. With this assumption, the channel capacity becomes: 

\begin{align}
C= \log_2 det(I_{n_{R}}+\frac{P}{\sigma^2  n_{R}} HH^{\dag}) \ \ \ bit/s/Hz
\end{align} \\

In the above formula, we have assumed that the available power at the transmit end is constant and equal all over the rays. \\
Nevertheless, we can define another setup in which the available power at the transmit end is equally distributed over rays. In this case, if the available power is $P$, each ray has the power of $\frac{P}{n_{R}}$.
We can therefore rewrite the formula we have for channel capacity for the second setup:
\begin{align}
C= \log_2 det(I_{n_{R}}+\frac{P}{\sigma^2  n_{R}  n_{T}} HH^{\dag}) \ \ \ bit/s/Hz
\end{align} \\

To recapitulate, the first and second setup follow formula [4] and [5], respectively. 

\subsection{Limit for Channel Capacity}
For the setup we defined in this paper, one can find the channel capacity when $n_{R}$ and  $n_{T}$ are large enough. \\
The channel capacity of our setup is saturated for large number of antennas and rays only if the transmit power is evenly distributed over the rays.\\
For the other setup in which the transmit power is constant and equal over the rays, there is no upper limit. Therefore, we allocated 1 section to find the channel capacity for the second setup when $n_{R}$ and  $n_{T}$ are large enough .
\subsubsection{Channel Capacity for the Second Setup}
When assuming the transmit power is evenly distributed over the rays, regardless of the number of rays, channel capacity follows formula [5]. As  $n_{T}$ gets large we have: \\
\begin{align}
\lim_{n_{T}\to\infty} HH^{\dag}={n_{T}} \times I_{n_{R}}
\end{align} \\
This can be proven by the law of large numbers and thus the capacity in the limit of large nT is:
\begin{align}
C= \log_2 det(I_{n_{R}}+\frac{P}{\sigma^2  n_{R}  n_{T}} \times  n_{T} \times I_{n_{R}})\\
=  \log_2 det(I_{n_{R}}+\frac{P}{\sigma^2  n_{R} } I_{n_{R}})\\
=\log_2 det(I_{n_{R}}(1+\frac{P}{\sigma^2  n_{R} } ))\\
=n_{R}\times \log_2 (1+\frac{P}{\sigma^2  n_{R} } )
\end{align} \\
But, what if both ${n_{R}}$ and  ${n_{T}}$ go to infinity? In this case, the formula derived above would be simplified to: \\

\begin{align}
C=\lim_{n_{R}\to\infty} n_{R}\times \log_2 (1+\frac{P}{\sigma^2  n_{R} })
=\lim_{x\to 0} \frac{\log_2 (1+\frac{xP}{\sigma^2})}{x}
\end{align}\\
where $x=\frac{1}{n_{R}}$.\\
\\By L'Hospital's rule we have: 
\begin{align}
\boxed{C=\frac{P}{\sigma^2 \times ln(2)}}
\end{align} \\
\section{Uniformly Spaced Planar Array}
We assume that the elements are arranged uniformly along a rectangular grid in yz-plane, with an element spacing $d_{y}$ in the y-direction and an element spacing $d_{z}$ in in the z-direction.
Since, the arrangement is Cartesian, it is useful to use two indices to refer to the elements: a row index and a column index. Grid indices in the y and z direction are denoted as m and n, respectively. The position vector of the $mn^{th}$ element is given by:

\begin{align}
\vec{r} _{mn}=x_{mn}\hat{x}+y_{mn}\hat{y}+z_{mn}\hat{z}
\end{align}
Assuming we have the spacing indicated, and the array starts at the origin, we can rewrite the position vector as follows:
\begin{align}
\vec{r} _{mn}=m\times d_{y}\hat{y}+m\times d_{z}\hat{z}
\end{align}
On the other hand, $\hat{r}$, a unit vector pointing in the direction of interest, can be written as: 
\begin{align}
 \hat{r}=sin\theta cos\phi \hat{x}+sin\theta sin\phi \hat{y}+cos\theta \hat{z}
\end{align}
For the linear configuration we had in the previous part, one can write: 
\\ $\vec{r}_{m}=md\hat{z}$, yielding: 
\begin{align}
\hat{r}.\vec{r}_{m}=(sin\theta cos\phi \hat{x}+sin\theta sin\phi \hat{y}+cos\theta \hat{z}). md\hat{z}=md_{z}cos\theta
\end{align}
, the formula we used above. 
However, for the planar configuration, this formula becomes: 
\begin{align}
\hat{r}.\vec{r}_{mn}=(sin\theta cos\phi \hat{x}+sin\theta sin\phi \hat{y}+cos\theta \hat{z}). (m\times d_{y}\hat{y}+m\times d_{z}\hat{z})=md_{y}sin\theta sin\phi+md_{z}cos\theta
\end{align}
We have two cases: 
\\
1. If the rays come from $\phi=0$, $sin\phi=0$ and the formula would be simplified to the same formula for a linear array. However, if the number of receive antenna in a square configuration is $n_{R}$, then the channel capacity would be the same for a linear array configuration with $\sqrt{n_{R}}$ antennas. 
\
This is intuitively correct and could be mathematically proven as well. 
\\
2. if the direction of incident ray make an angle $\phi \neq 0$ with configuration plane. 
In this case,  the channel capacity slightly changes for different incident angles; nevertheless, the amount of variation is not significant. 
To recapitulate, with the same number of antenna elements, linear array always outperforms square (or rectangular) array in terms of channel capacity. 
\\ The simulation results for planer array discussed here are shown in the second part. 
\section{Placing satellites around the Earth}
In this section, we assume that we want to put some satellites around the Earth serving terrestrial users such that the amount of interference is minimized. To do so, firstly, we put satellites so that the minimum distance between them (considering all possible pairs) is maximized. This question intuitively reminds us of the “Tammes problem” which has been extensively discussed before. \\
Tammes [5] problem looks for an answer for the following question: “How must N congruent non-overlapping spherical caps be packed on the surface of a unit sphere so that the angular diameter of spherical caps will be as great as possible”\\
In the above statement, the circle on the surface of a sphere is called a spherical cap. \\
One can easily correspond our problem to the answer of Tammes problem. We can say all satellites are located on a unique sphere when revolving around the Earth. It is worth mentioning to note that when a satellite is relatively close to Earth, the orbit on which the satellite traverses on is roughly a circle, and therefore our assumption is valid.\\
 Tammes problem was solved for some specific number of points(for N=1,2,…,12,23,24 and some other values).\\
Let X be a finite subset of $S^{n-1}$ in $\mathbb{R}^{n}$. We define $\psi$ as follows \cite{mohajer2018new}:
\begin{align}
\psi(x)=\min_{x,y \in X} {dist(x,y)},   x\neq y
\end{align}
Then X is a spherical $\psi(X)$-code. Also, define $d_{N}$ the largest angular separation $\psi(X)$ with $|X|=N$ that could be obtained in $S^2$, meaning that:
\begin{align}
d_{N}=\max_{X \subset S^2} {\psi (X)}, |X|=N.
\end{align}
\\ In the following table, $d_{N}$ is shown for some N (the values were found by different persons at different times):
\\
\begin{tabular}{ |p{3cm}||p{3cm}|  }
 \hline
 \multicolumn{2}{|c|}{Largest Angular Separation for Different Values of N} \\
 \hline
 N&d\\
 \hline
 4   &  109.4712206\\
5  &  90.0000000\\
6  &  90.0000000\\
7  & 77.8695421\\
8  & 74.8584922\\
9  & 70.5287794\\
10  & 66.1468220\\
11  & 63.4349488\\
12  & 63.4349488\\
13  & 57.1367031\\
14  & 55.6705700\\
15 & 53.6578501\\
16 & 52.2443957\\
17 & 51.0903285\\
 \hline
\end{tabular}
\subsection{No interference}
In this part, we assume that each non-overlapping spherical cap found for each N is the terrestrial coverage area for the corresponding satellite. In this case, each server on Earth is served by a single satellite, and therefore there is no interference. Having this setup, there exist some place on Earth not being served by any satellites. For this matter, we define coverage percentage for each configuration as ratio of the area covered by satellites to surface area of Earth. 
\\
The following table could be accordingly attained.  
\\
\\
\begin{tabular}{ |p{3cm}||p{3cm}|  }
 \hline
 \multicolumn{2}{|c|}{Coverage Percentage for Different Values of N } \\
 \hline
 N&Coverage Percentage\\
 \hline
 4   &  0.8386\\
5  &  0.7322\\
6  &  0.8787\\
7  & 0.7775\\
8  & 0.8234\\
9  & 0.8258\\
10  & 0.8101\\
11  & 0.8214\\
12  & 0.8961\\
13  & 0.7914\\
14  & 0.8099\\
15 & 0.8073\\
16 & 0.8171\\
17 & 0.8309\\
 \hline
\end{tabular}
\\
\\
As seen, the coverage percentage is non-linear as N grows, however, it reaches it maximum value for N=12 among the values considered above. 
\subsection{Interference and Overlapping Coverage Area}
If one need to cover the whole surface of Earth with existing \cite{maurer1999unconditionally,maurer1993secret,wilson2007channel,ye2007secrecy} satellites, the coverage area for each satellite needs to be enlarged. For this purpose, each spherical cap is equally enlarged till all points on the surface of the Earth would be covered by at least on satellite.
\\ 
In this case, there would be some terrestrial servers receiving signals from a couple of satellites. From [6], for $N>6$, if a server receives signal from more than 1 satellites, the number of satellites seen by the server is at most 5 and at least 3. 
\\
In [7], the author tries to find conjectured solutions for this problem. Also, it defined density denoted by $D_{N}$ which has the same meaning as Coverage Percentage defined in this paper. 
\\
Moreover, $r_{N}$ represents angular radius and in the following table the maximum value of $r_{N}$ for each configuration is written.
\\
\\
\begin{tabular}{ |p{3cm}||p{3cm}||p{3cm}|  }
 \hline
 \multicolumn{3}{|c|}{Conjectured Covering of a Sphere by N equal circles } \\
 \hline
 N& $r_{N}$ & Coverage Percentage\\
 \hline
 4   & 70.5287 & 1.3333\\
5  &  63.4349 & 1.3819\\
6  &  54.7356 & 1.2679\\
7  & 51.0265 & 1.2986\\
8  & 48.1395 & 1.3307\\
9  & 45.8788 & 1.3672\\
10  & 42.3078 & 1.3023\\
11  & 41.4271 & 1.3761\\
12  & 37.3773 & 1.2320\\
13  & 37.0685 & 1.3135\\
14  & 34.9379 & 1.2615\\
15 & 34.0399 & 1.2851\\ 
16 & 32.8988 & 1.2829\\
17 & 32.0929 & 1.2989\\
 \hline
\end{tabular}

\section{Simulation Results}
Simulations have been done in MATLAB for both setups. For each setup, channel capacity is depicted versus $n_{R}$ when $n_{T}$ is constant, also versus $n_{T}$ when $n_{R}$ is constant. For the second setup in which there exists a upper bound for channel capacity, the simulation results have been also compared with the results we found in the previous section. 
\subsection{First setup}
When we assume the transmit power is constant and equal over each rays. \\
We consider two cases:
First, when the number antennas varies from 1 to 50, while the number of rays is constant. 
\begin{figure}
\centering
  \includegraphics[width=0.5\linewidth]{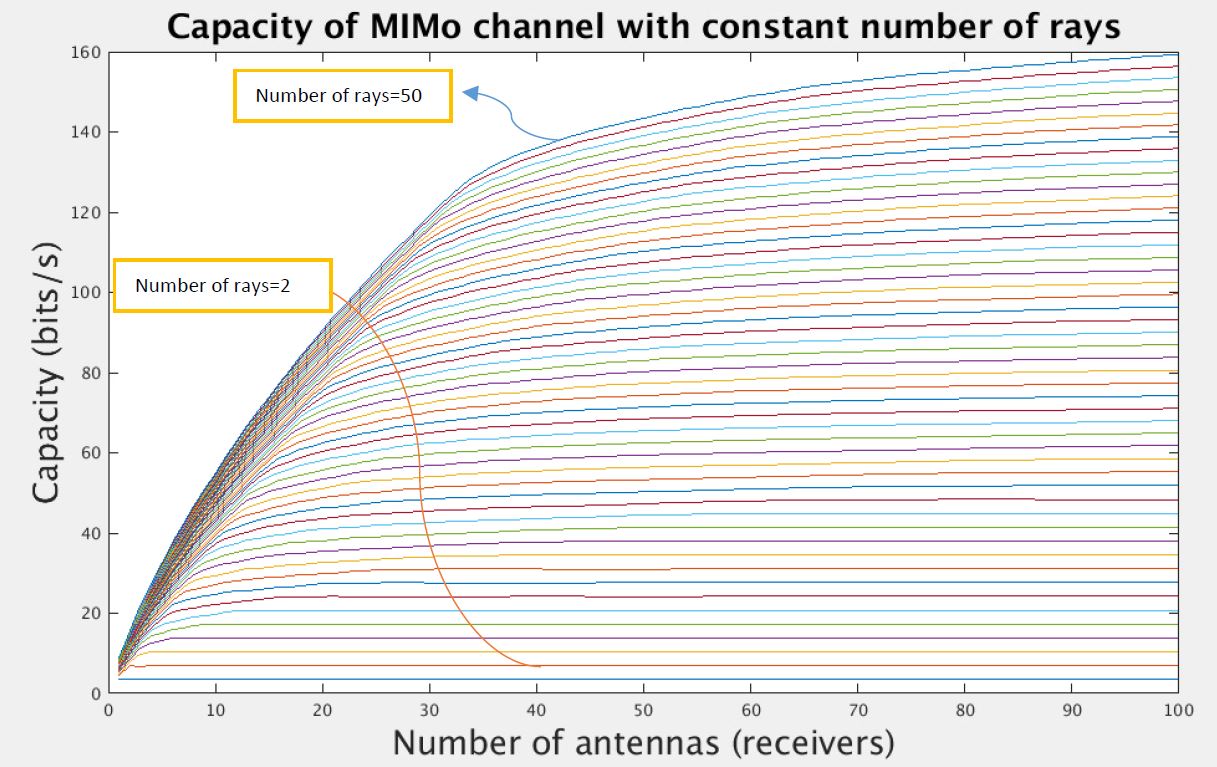}
 \caption{Channel Capacity of MIMO channel, when the number antennas varies from 1 to 50, while the number of rays is constant }
\end{figure}

Second, when the number rays varies from 1 to 50, while the number of antennas is constant. 
\begin{figure}
\centering
  \includegraphics[width=0.5\linewidth]{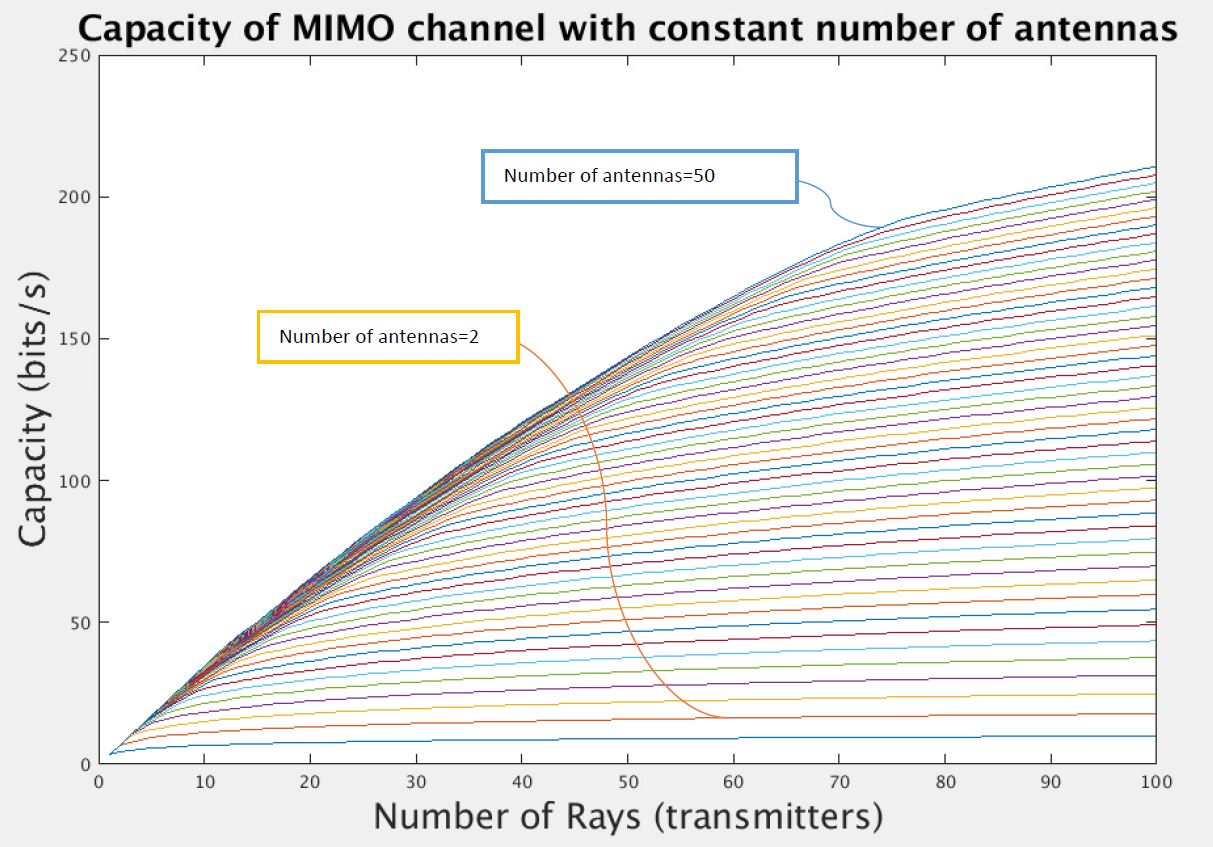}
 \caption{Channel Capacity of MIMO channel, when the number rays varies from 1 to 50, while the number of antennas is constant. }
\end{figure}
\subsection{Second setup}
When we assume the transmit power is evenly distributed over the rays.
We consider two cases:
First, when the number antennas varies from 1 to 50, while the number of rays is constant. 
\begin{figure}
\centering
  \includegraphics[width=0.5\linewidth]{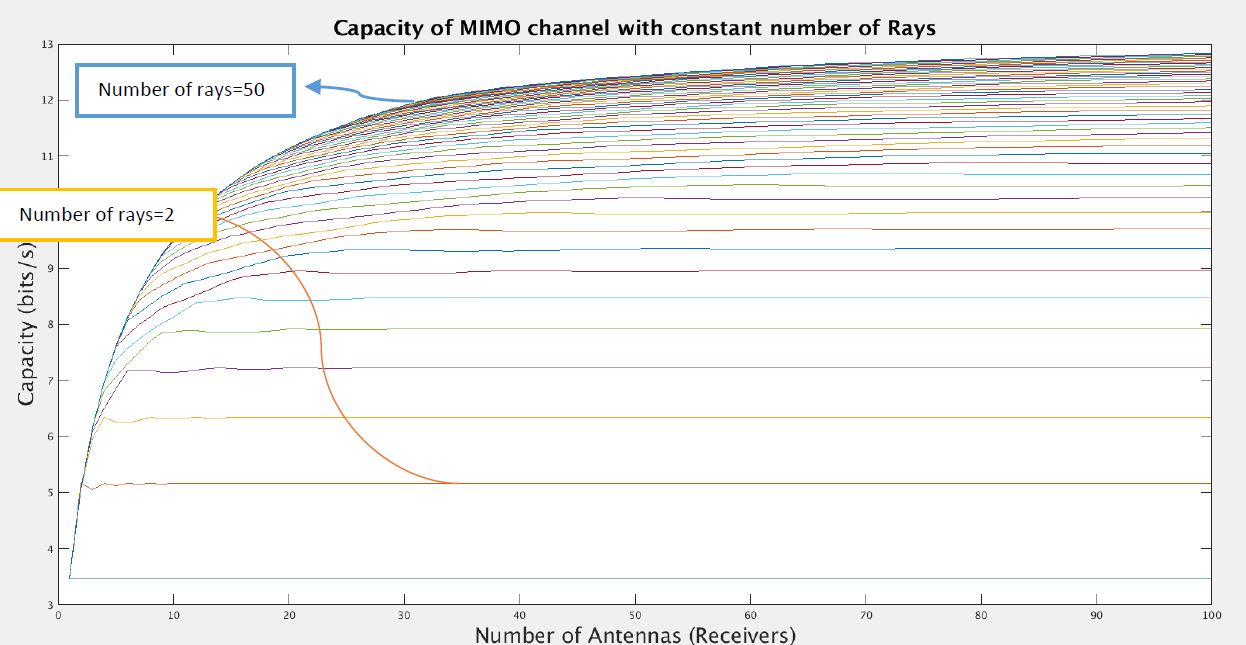}
 \caption{Channel Capacity of MIMO channel, when the number antennas varies from 1 to 50, while the number of rays is constant }
\end{figure}

Second, when the number rays varies from 1 to 50, while the number of antennas is constant. 
\begin{figure}
\centering
  \includegraphics[width=0.5\linewidth]{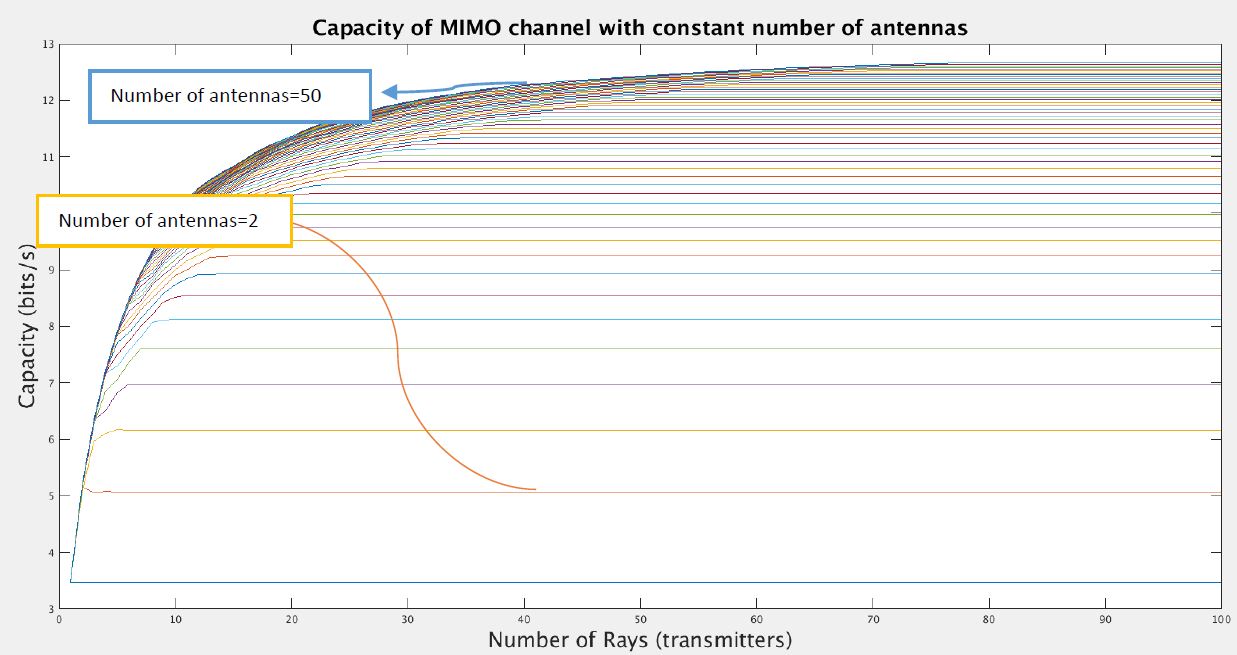}
 \caption{Channel Capacity of MIMO channel, when the number rays varies from 1 to 50, while the number of antennas is constant. }
\end{figure}
\subsection{Comparing with AWGN channel}
Similar to the previous part, we will consider both setups. SISO AWGN channel is compared with both setups when: $n_{R},n_{T} \in {10,20}$

\begin{figure*}
  \centering
  \begin{subfigure}[b]{0.39\linewidth}
    \includegraphics[width=\linewidth]{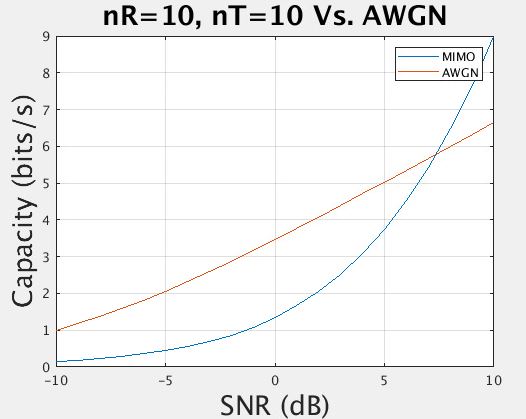}
    \caption{ $n_{R}=10$  and $n_{T}=10$ }
  \end{subfigure}
  \begin{subfigure}[b]{0.4\linewidth}
    \includegraphics[width=\linewidth]{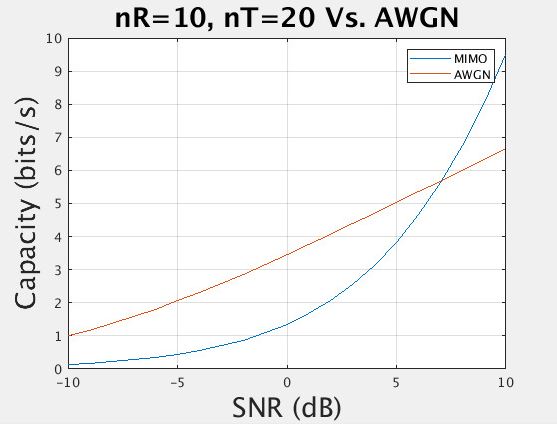}
    \caption{ $n_{R}=10$  and $n_{T}=20$ }
  \end{subfigure}
\begin{subfigure}[b]{0.4\linewidth}
    \includegraphics[width=\linewidth]{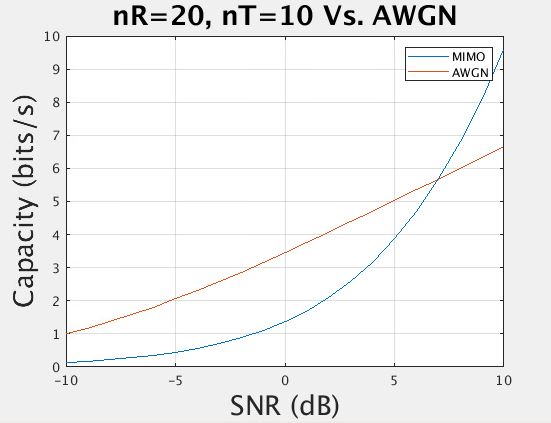}
    \caption{$n_{R}=20$  and $n_{T}=10$ }
  \end{subfigure}
  \begin{subfigure}[b]{0.4\linewidth}
    \includegraphics[width=\linewidth]{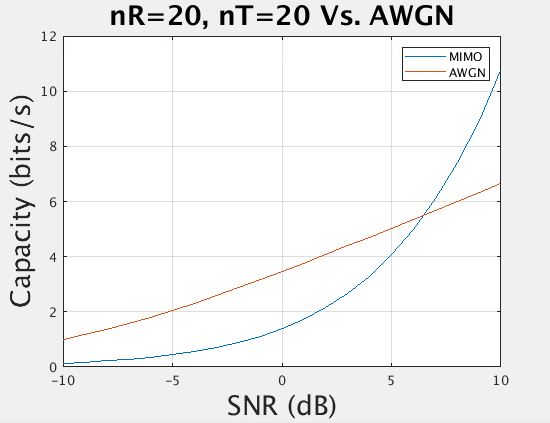}
    \caption{ $n_{R}=20$  and $n_{T}=20$ }
  \end{subfigure}
  \caption{Comparison between Channel capacity of MIMO setup (first setup) defined in this paper and legacy SISO AWGN channel }
\end{figure*}

\begin{figure*}
  \centering
  \begin{subfigure}[b]{0.39\linewidth}
    \includegraphics[width=\linewidth]{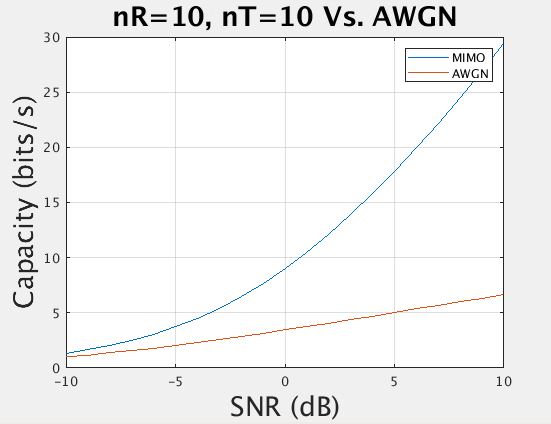}
    \caption{ $n_{R}=10$  and $n_{T}=10$ }
  \end{subfigure}
  \begin{subfigure}[b]{0.4\linewidth}
    \includegraphics[width=\linewidth]{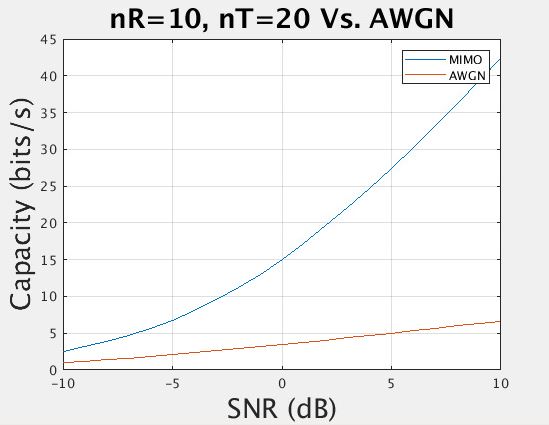}
    \caption{ $n_{R}=10$  and $n_{T}=20$ }
  \end{subfigure}
\begin{subfigure}[b]{0.4\linewidth}
    \includegraphics[width=\linewidth]{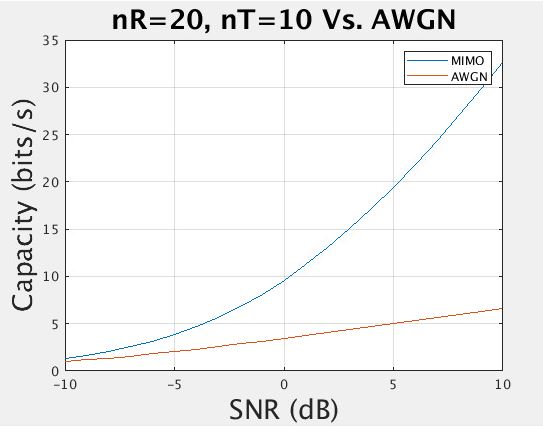}
    \caption{$n_{R}=20$  and $n_{T}=10$ }
  \end{subfigure}
  \begin{subfigure}[b]{0.4\linewidth}
    \includegraphics[width=\linewidth]{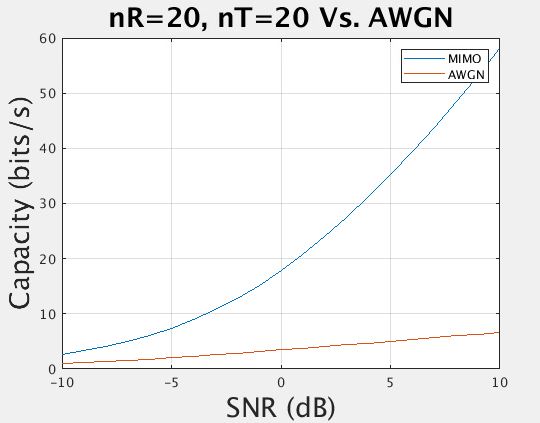}
    \caption{ $n_{R}=20$  and $n_{T}=20$ }
  \end{subfigure}
  \caption{Comparison between Channel capacity of MIMO setup  (Second setup) defined in this paper and legacy SISO AWGN channel }
\end{figure*}
\subsection{Square configuration} 
Firstly, we evaluate the effect of $\phi$ in channel capacity. To do so, 4 different scenarios have been simulated here:
\begin{itemize}
\item $n_{T}=1...20, n_{R}=16 (4\times4 square)$
\item $n_{T}=1...20, n_{R}=25 (5\times5 square)$
\item $n_{T}=1...20, n_{R}=36 (6\times6 square)$
\item $n_{T}=1...20, n_{R}=49 (7\times7 square)$
\end{itemize}
\begin{figure*}
  \centering
  \begin{subfigure}[b]{0.39\linewidth}
    \includegraphics[width=\linewidth]{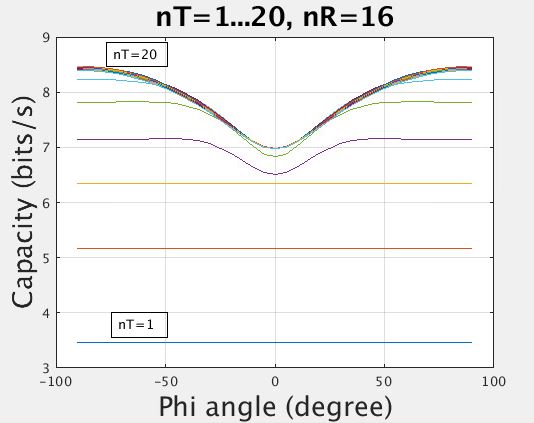}
    \caption{ $n_{R}$ varies from 1 to 20  and $n_{T}=16$ }
  \end{subfigure}
  \begin{subfigure}[b]{0.4\linewidth}
    \includegraphics[width=\linewidth]{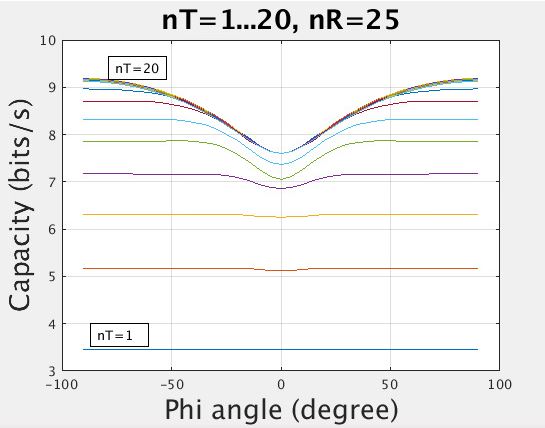}
    \caption{ $n_{R}$ varies from 1 to 20  and $n_{T}=25$ }
  \end{subfigure}
\begin{subfigure}[b]{0.4\linewidth}
    \includegraphics[width=\linewidth]{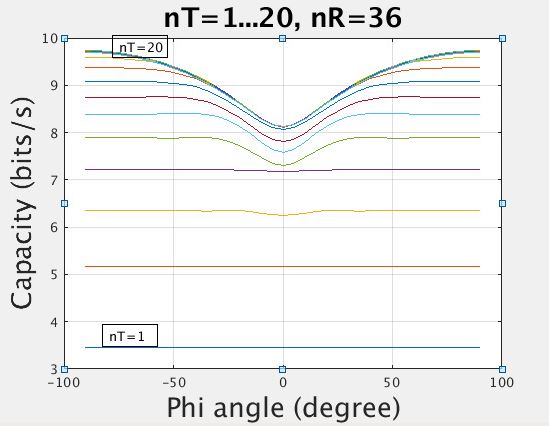}
    \caption{ $n_{R}$ varies from 1 to 20  and $n_{T}=36$ }
  \end{subfigure}
  \begin{subfigure}[b]{0.4\linewidth}
    \includegraphics[width=\linewidth]{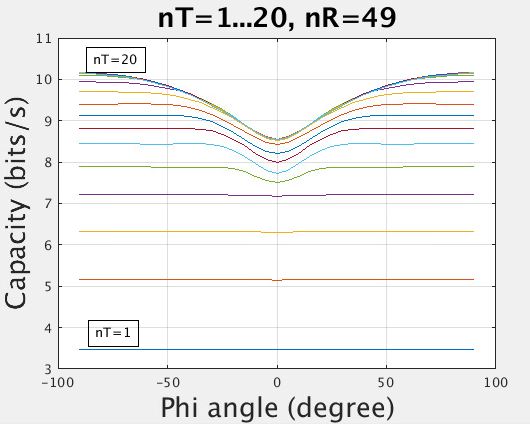}
    \caption{ $n_{R}$ varies from 1 to 20  and $n_{T}=49$ }
  \end{subfigure}
  \caption{Effect of $\phi$ on channel capacity for 4 different square configurations}
\end{figure*}
These values show that the maximum value of channel capacity occurs at $\phi=90^\circ$ no matter what $nT$ and $nR$ are. 
\\
Secondly, square and linear configurations would be compared.\\
Henceforth, we consider  $\phi=90^\circ$ to compare square and linear configurations. We have simulated 4 cases: 

\begin{itemize}
\item $n_{T}=10, n_{R}=16 (4\times4 square)$
\item $n_{T}=20, n_{R}=16 (4\times4 square)$
\item $n_{T}=10, n_{R}=25 (4\times4 square)$
\item $n_{T}=20, n_{R}=25 (4\times4 square)$
\end{itemize}

\begin{figure*}
  \centering
  \begin{subfigure}[b]{0.39\linewidth}
    \includegraphics[width=\linewidth]{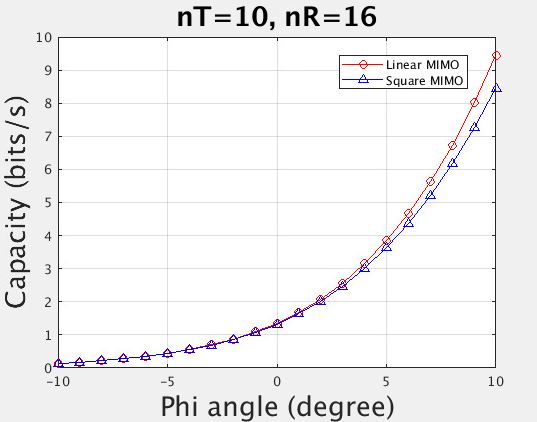}
    \caption{ $n_{R}=10$  and $n_{T}=16$ }
  \end{subfigure}
  \begin{subfigure}[b]{0.4\linewidth}
    \includegraphics[width=\linewidth]{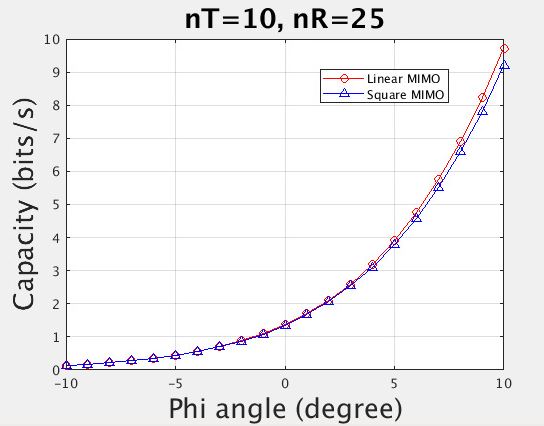}
    \caption{ $n_{R}=10$  and $n_{T}=25$ }
  \end{subfigure}
\begin{subfigure}[b]{0.4\linewidth}
    \includegraphics[width=\linewidth]{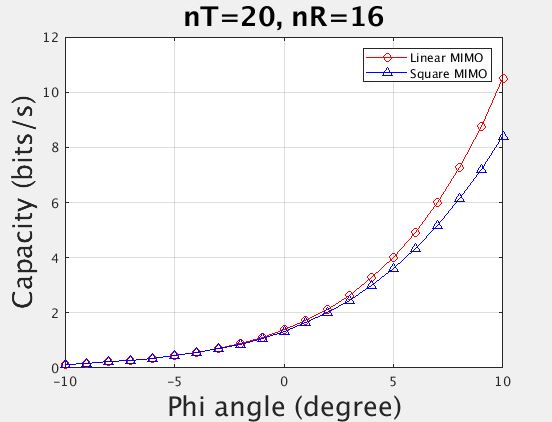}
    \caption{ $n_{R}=20$  and $n_{T}=16$ }
  \end{subfigure}
  \begin{subfigure}[b]{0.4\linewidth}
    \includegraphics[width=\linewidth]{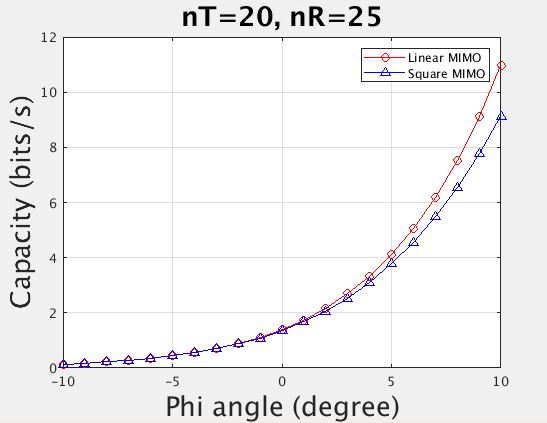}
    \caption{ $n_{R}=20$  and $n_{T}=25$ }
  \end{subfigure}
  \caption{Linear Vs. Square configuration with the same number of elements. $\phi$ is constant and equal to $90^\circ$ to maximize the capacity}
\end{figure*}

\bibliographystyle{IEEEtran}
\bibliography{refs}

\end{document}